\def\ts     {\thinspace}
\def\kms    {\ifmmode{{\rm \ts km\ts s}^{-1}}\else{\ts km\ts s$^{-1}$}\fi}
\def\msol   {\ifmmode{{\rm M}_{\odot} }\else{M$_{\odot}$}\fi}
\def\lsol   {\ifmmode{L_{\odot}}\else{$L_{\odot}$}\fi}
\def\lfir   {\ifmmode{L_{\rm FIR}}\else{$L_{\rm FIR}$}\fi}
\def\zsol   {\ifmmode{{\rm Z}_{\odot}}\else{Z$_{\odot}$}\fi}
\def\etal   {{\rm et\ts al.}}
\def\aco    {\ifmmode{{\rm CO}(J\!=\!1\! \to \!0)}\else{{\rm CO}($J$=1$\to$0)}\fi}
\def\bco    {\ifmmode{{\rm CO}(J\!=\!2\! \to \!1)}\else{{\rm CO}($J$=2$\to$1)}\fi}
\def\cco    {\ifmmode{{\rm CO}(J\!=\!3\! \to \!2)}\else{{\rm CO}($J$=3$\to$2)}\fi}
\def\dco    {\ifmmode{{\rm CO}(J\!=\!4\! \to \!3)}\else{{\rm CO}($J$=4$\to$3)}\fi}
\def\eco    {\ifmmode{{\rm CO}(J\!=\!5\! \to \!4)}\else{{\rm CO}($J$=5$\to$4)}\fi}
\def\fco    {\ifmmode{{\rm CO}(J\!=\!6\! \to \!5)}\else{{\rm CO}($J$=6$\to$5)}\fi}
\def\gco    {\ifmmode{{\rm CO}(J\!=\!7\! \to \!6)}\else{{\rm CO}($J$=7$\to$6)}\fi}
\def\hco    {\ifmmode{{\rm CO}(J\!=\!8\! \to \!7)}\else{{\rm CO}($J$=8$\to$7)}\fi}
\def\ico    {\ifmmode{{\rm CO}(J\!=\!9\! \to \!8)}\else{{\rm CO}($J$=9$\to$8)}\fi}
\def\jco    {\ifmmode{{\rm CO}(J\!=\!10\! \to \!9)}\else{{\rm CO}($J$=10$\to$9)}\fi}
\def\kco    {\ifmmode{{\rm CO}(J\!=\!11\! \to \!10)}\else{{\rm CO}($J$=11$\to$10)}\fi}
\def\ci     {\ifmmode{{\rm C}{\rm \small I}}\else{C\ts {\scriptsize I}}\fi}
\def\hi     {\ifmmode{{\rm H}{\rm \small I}}\else{H\ts {\scriptsize I}}\fi}
\def\hh     {\ifmmode{{\rm H}_2}\else{H$_2$}\fi}
\def\cone {\ifmmode{{\rm C}{\rm \small I}(^3\!P_1\!\to^3\!P_0)}
     \else{C\ts {\scriptsize I}{\small$(^3\!P_1\!\to^3\!\!\!P_0)$}}\fi}
\def\ctwo {\ifmmode{{\rm C}{\rm \small I}(^3\!P_2\!\to^3\!P_1)}
     \else{C\ts {\scriptsize I}{\small$(^3\!P_2\!\to^3\!\!\!P_1)$}}\fi}
\def\cij {\ifmmode{{\rm C}{\rm \small I}\,(^3P_i\to^3P_j)}\else{C\ts {\scriptsize I}\,{\small$(^3P_i\to^3P_j)$}}\fi}
\def\cii    {\ifmmode{{\rm C}{\rm \small II}}\else{C\ts {\scriptsize II}}\fi}
\def\tex {\ifmmode{{T}_{\rm ex}}\else{$T_{\rm ex}$}\fi}
\def\tmb {\ifmmode{{T}_{\rm mb}}\else{$T_{\rm mb}$}\fi}
\def\tkin {\ifmmode{{T}_{\rm kin}}\else{$T_{\rm kin}$}\fi}
\def\microns {\ifmmode{\mu{\rm m}}\else{$\mu$m}\fi}
\def\um{\ifmmode{\mu{\rm m}}\else{$\mu$m}\fi}
\def\nhh   {\ifmmode{n({\rm H}_2)}\else{$n$(H$_2$)}\fi}
\def\gradv {\ifmmode{{\rm dv/dr}}\else{dv/dr}\fi}
\documentclass[twocolumn]{aa}
\usepackage{graphicx}
\begin{document}
 \title{LABOCA observations of nearby, active galaxies}

   \author{A. Wei\ss
          \inst{}
          \and
          A. Kov\'{a}cs
          \inst{}
          \and
          R. G\"usten
          \inst{}
          \and
          K. M. Menten
          \inst{}
          \and 
         F. Schuller
          \inst{}
          \and
          G. Siringo
          \inst{}
	 \and
          E. Kreysa
          \inst{}		
          }

   \institute{{ Max-Planck-Institut f\"ur Radioastronomie}, Auf dem H\"ugel 69, 53121 Bonn, Germany
             }

   \date{}

   \abstract{We present large scale 870\,\um\ maps { of} the nearby starburst galaxies NGC\,253 and NGC\,4945 as well as the 
nearest giant elliptical radio galaxy Centaurus A (NGC\,5128) obtained with the newly commissioned {Large Apex Bolometer Camera (LABOCA)} operated at 
the { Atacama Pathfinder Experiment} telescope. Our continuum images reveal for the first time 
the distribution of cold dust at a angular resolution of { 20$''$} across the entire optical disks of NGC\,253 
and NGC\,4945 out to a radial distance of $10'$ (7.5 kpc). In NGC\,5128 our LABOCA image also shows, for 
the first time { at submillimeter wavelengths}, the synchrotron emission associated with the radio jet and the inner radio lobes.
From an analysis of the 870\,\um\ emission in conjunction with ISO-LWS, IRAS and long wavelengths radio data 
we find temperatures for the cold dust in the disks of all three galaxies of 17--20\,K, comparable to the dust 
temperatures in the disk of the Milky Way. The total gas mass in the three galaxies 
is determined to be 2.1, 4.2 and $2.8\times10^{9}\,\msol$ for NGC\,253, NGC\,4945 and NGC\,5128, respectively. 
The mass of the warmer (30--40\,K) gas associated with the central starburst regions in NGC\,253 and NGC\,4945 
only accounts for $\sim10\%$ of the total gas mass. A detailed comparison between the gas masses derived from the dust 
continuum and the integrated CO(1--0) intensity in NGC\,253 suggests that changes of the CO luminosity to molecular
mass conversion factor are mainly driven by a metallicity gradient and only to a lesser degree by variations of the 
CO excitation. An analysis of the synchrotron spectrum in the northern radio lobe of NGC\,5128 shows that the 
synchrotron emission from radio to the ultraviolet (UV) wavelengths is well described by a broken power law and that the break frequency is a 
function of the distance from the radio core as expected for aging electrons. We derive an outflow speed of 
$\sim0.5$\,{\it c} at a distance of 2.6\,kpc from the center, consistent with the speed derived in the vicinity of the nucleus.

   }

   \maketitle
%

\section{Introduction} 
Nearly half the bolometric luminosity in the local universe is emitted at mid- and far-infrared (IR) wavelengths. The 
IR radiation is produced by warm interstellar dust grains that are heated by UV photons from hot high mass stars. This 
thermal emission of dust grains therefore carries valuable information on feedback processes from star formation, the chemical 
composition of the { interstellar medium (ISM)} and also on the total amount of interstellar matter in galaxies, 
the gas surface density and its relation to star-forming regions.
The high spatial resolution of the Spitzer Space Telescope has greatly improved our view in this important { wavelength} range and allows for
the first time to study spatially resolved spectral energy distributions (SEDs) across galaxies and { the relation of IR emission} 
to other tracers of star formation. \\
In order to provide a complete measurement of the dust SED it is highly desirable to combine the IR observations with data on the long wavelength
(Rayleigh-Jeans) tail of the SED in the submm regime (see, e.g., Draine \etal\ \cite{draine07}). This is because the submm observations are
particularly sensitive to cold dust which dominates the dust mass in galaxies. \\
Ground based observations of the submm emission of nearby galaxies, however, remain a challenging task because they largely 
suffer from limitations due to the Earth's atmosphere. Furthermore, existing submm cameras have been limited in their field of view (FoV) to
a few arc minutes which makes it difficult to survey large areas on the sky in a reasonable amount of time.\\
With the commissioning of the Large APEX Bolometer Camera (LABOCA, Siringo \etal\ \cite{siringo07} and in prep.) 
at the APEX telescope  (G\"usten \etal\ \cite{guesten06}) at 5100\,m altitude on Chajnantor this situation 
has largely improved. With its large field of view and large number of detectors in combination with the extremely dry conditions at the 
site, LABOCA provides the mapping-speed and sensitivity required to survey large areas on the sky down to mJy noise levels.\\
In this paper we present the first large scale 870\,\um\ maps { by LABOCA} towards two
nearby starburst galaxies NGC\,253 and NGC\,4945, and towards the nearest giant elliptical radio galaxy Cen\,A (NGC\,5128). 
In Sect.\,2 we describe the LABOCA observations and the data reduction, Sect.\,3 focuses on the dust SEDs and 
gas masses. In Sect.\,4 we discuss the thermal and non-thermal emission processes in our target galaxies and Sect.\,5 summarizes our results.

\section{Observations and data reduction}
\begin{figure} 
\centering
\includegraphics[width=8.5cm,angle=0]{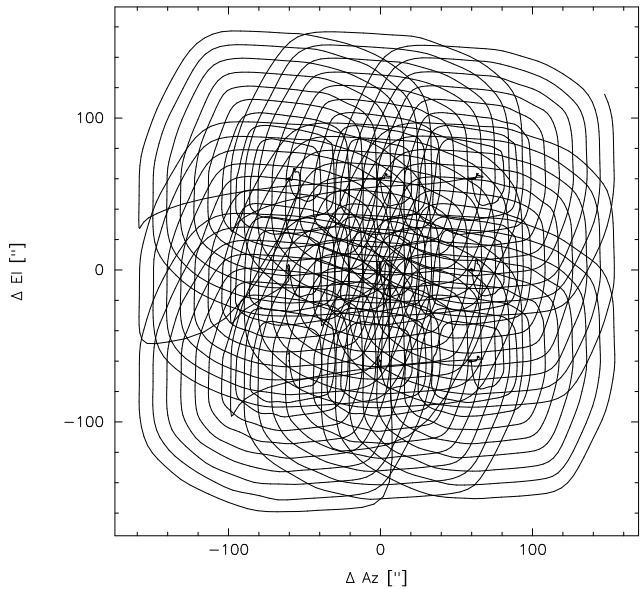} 
\caption{{ Azimuth-Elevation path of the central bolometer of the array for a single raster of spiral scan as used for the observations.}}
\label{spiral_image} 
\end{figure}

\noindent Observations were carried out using LABOCA on APEX. LABOCA is an array of 295 composite bolometers
with neutron-transmutation-doped (NTD) germanium thermistors. The bolometers are AC-biased and operated in total power mode.
Real-time signal processing of the 1\,KHz data stream includes digital anti-alias filtering and 
down-sampling to 25\,Hz. The radiation is coupled onto the detectors through an array of conical 
feed horns whose layout leads to a double beam spaced distribution of the individual beams in a hexagonal 
configuration over the $11'.4$ field of view. The center frequency of LABOCA is 345\,GHz and its passband has a 
FWHM of $\sim60$\,GHz. The measured angular resolution of each beam is $19''.2$.\\
\noindent The observations were carried out in 2007 July and August in mostly
excellent weather conditions (precipitable water vapor (PWV) typically 0.5 mm, corresponding to a zenith opacity
of 0.2 at the observing wavelength). Mapping was performed using a raster of spirals pattern. { In each scan,} 
the telescope traces { a set of} spirals with radii between { $20''$ and $1.6'$ at nine 
raster positions separated by 60$''$} in azimuth and elevation (see Fig.\,\ref{spiral_image}). 
{ The pattern leads} to a fully sampled map of the LABOCA FoV in a single scan. 
{ The radii and spacings of the spirals} were optimized { for} uniform { noise coverage} across the FoV{ , 
while keeping telescope overheads at a minimum.} { The scanning speed varies between 0.5'-- 2.5'\,s$^{-1}$, modulating the
source signals even from a wide range of scales into the useful post-detection frequency band (0.1 to 12.5 Hz) of LABOCA, 
while providing at least 3 measurements per beam at the data rate of 25 samples per second even at the highest scanning 
velocity.}

\noindent Calibration was achieved through observations of Mars, Uranus and Neptune as well as secondary calibrators
and was found to be accurate within { 9\% rms}. The atmospheric attenuation was determined via skydips every $\sim$ 2 hours
as well as from independent data from the APEX radiometer which measures the line of sight water vapor column every minute. 
{ The zenith opacities determined from both methods correlate well in general but the radiometer opacities are on average 30\% higher than the 
values derived from the skydips. Details on the opacity determination and its limitations will be given by Siringo \etal\ (in prep).
We used a linear combination of both values for the data described here, with weights chosen such that they provide the most consistent 
calibrator fluxes for a wide range of opacities and source elevations.}\\
Focus settings were { determined typically} once per night and checked during sunrise. Pointing was checked on 
nearby quasars and found to be stable within $3''$ { rms}.\\
\noindent { As all of our target galaxies are larger than the FoV of LABOCA, maps centered at several pointings were made towards each
galaxy.} The final maps consist of 3 pointings for NGC\,253 and NGC\,4945 and 5 pointings for Cen\,A. The total
on source observing time is 8h, 3h and 5h for NGC\,253, NGC\,4945 and Cen\,A respectively.\\
\noindent The data was reduced using the BoA reduction package (Schuller \etal\ in prep.). Reduction steps on the time 
series {(time ordered data of each bolometer)} include temperature drift correction based on two "blind" bolometers 
(whose horns have been sealed to block the sky signal), flat fielding, calibration, opacity correction, correlated noise 
removal on the full array as well as on groups of bolometers related by the wiring and in the electronics, flagging of 
{ unsuitable data (bad bolometers and/or data taken outside reasonable telescope scanning velocity and acceleration 
limits) as well as} de-spiking. Each reduced scan was then gridded into a spatial intensity and weighting map. Weights are 
calculated based on the rms of each time series contributing to a certain grid point in the map. Individual maps were 
co-added noise-weighted and finally smoothed to an angular resolution of $20''$.
The resulting rms { noise level} for the central part of the maps is 3.5, 5.5 and 4.0 mJy/beam for  NGC\,253, NGC\,4945 and Cen\,A, 
respectively. The final maps are shown in Figs.\,\ref{n253_image},\,\ref{n4945_image}\,and\,\ref{cena_image}.\\
{ The correlated noise removal implies that structures of a size similar to the angular separation on the sky between 
bolometer involved in the de-correlation will be removed from our maps (for details see e.g. Kov\'{a}cs \etal\ \cite{kovacs08}). 
This effect is most pronounced for the correlated noise removal based on the wiring which typically involves 20 bolometers
covering angular scales of $5'\times 1.5'$ on sky. As a result, some of the faint, extended emission may have been 
filtered in our reduction. However, owing to the substantial field rotation (vs. the array alignment) over the span of the 
observations, we expect to fully recover scales up to 5' along all spatial directions in our reductions. Since all of our targets 
are close to edge--on systems, with minor axes below this limiting scale, most of the large scale emission should be adequately 
recovered in the analysis. }

\section{Results}
\subsection{NGC\,253}
\begin{figure*} 
\centering
\includegraphics[width=16.5cm,angle=0]{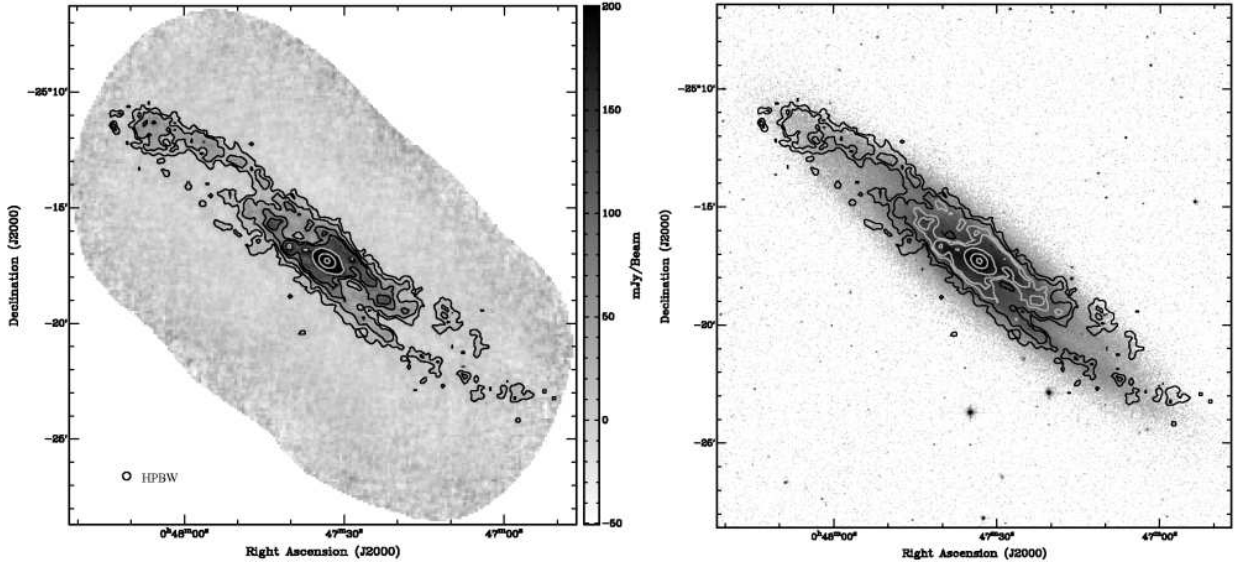} 
\caption{{\it Left:} LABOCA 870\,\um\ flux density map of NGC\,253 smoothed to a spatial resolution of 20$''$. 
Contours are shown for 0.01, 0.03, 0.07, 0.1 (black) and 0.2, 1.0, 3.0 Jy/beam (grey). {\it Right:} 870\,\um\ contours 
overlaid onto a 2MASS K-band image of NGC\,253. Contour levels are the same as in the left part of the figure { but only the two lowest
levels are shown in black.}}
\label{n253_image} 
\end{figure*}

\begin{figure*} 
\centering
\includegraphics[width=16.5cm,angle=0]{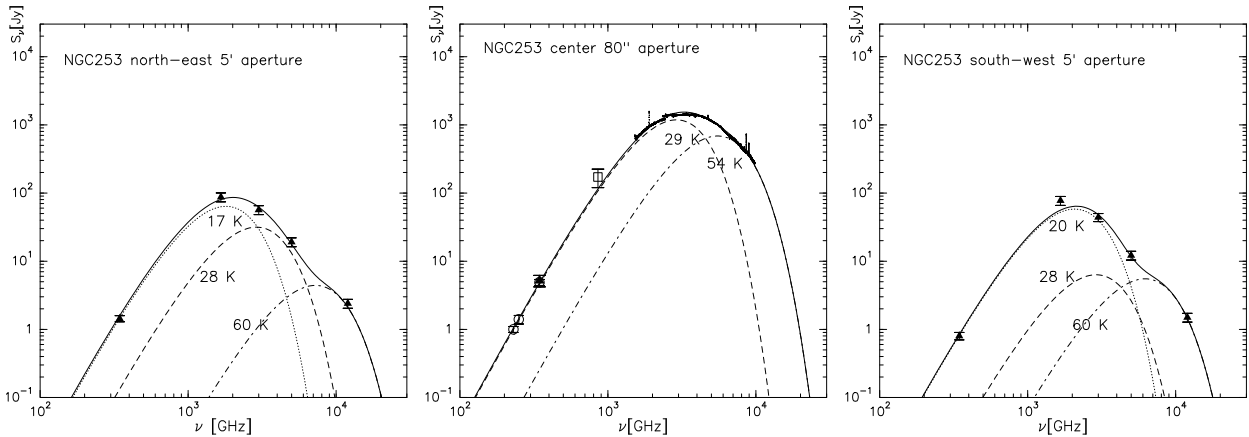} 
\caption{Spectral energy distributions towards the north-east (left), center (middle) and south-west (right) of NGC\,\,253.
The 180, 100, 60 and 25\,\um\ data points in the NE and SW are from Radovich \etal\ \cite{radovich01}. The { 1.3mm} and 350\,\um\ data in
the center are from Kr\"ugel \etal\ (\cite{kruegel90}) and Rieke \etal\ (\cite{rieke73})}
\label{n253_dust} 
\end{figure*}

{\it 870\,\um\ morphology:} The 870\,\um\ distribution in the central region of NGC\,253 is dominated by the compact, well studied  
nuclear starburst region (see Fig\,\ref{n253_image}). { Fitting a two-dimensional Gaussian intensity distribution yields a deconvolved 
size} (FWHM) of $30''\times16''$ ($370\times200$\,pc at an assumed distance of 2.5\,Mpc, Houghton \etal\ \cite{houghton97}) 
and a total flux is $6.4\pm0.7$ Jy. 
The dust surrounding the central starburst region closely follows the stellar bar visible at near infrared wavelengths (for an overlay
of the 870\,\um\ emission onto a NIR K band image see Fig.\,\ref{n253_image} right). The dust distribution in this inner part
of NGC\,253 is consistent with previously published Submillimeter Common User Bolometer Array (SCUBA) maps 
(Alton \etal\ \cite{alton99}). The large scale dust distribution
is dominated by two prominent spiral arms which originate near the stellar bar and are detected throughout the optical
disk of NGC\,253 out to a distance of $10'$ from the nucleus. More diffuse emission is detected towards the trailing edge of 
the disk. Our LABOCA map does not show significant dust continuum emission associated with the outflow along the minor axis of 
NGC\,253 (e.g. McCarthy \etal\ \cite{mcc87}, Fabbiano \cite{fabbiano88}, Dahlem \etal\ \cite{dahlem98}). 
The total 870\,\um\ flux detected towards NGC\,253 is $17.6\pm1.8$ Jy.\\
{\it Dust temperature and mass:} The dust temperature distribution in NGC\,253 has been investigated by 
Radovich \etal\ \cite{radovich01} and Melo \etal\ \cite{melo02} based on ISOPHOT and IRAS mid and far-IR observations. 
Their analysis, however, does not include data on the Rayleigh-Jeans tail of the dust spectrum (except for the 
nuclear region of NGC\,253) which could potentially hamper the identification of cold dust.\\
We have reanalyzed the dust temperatures in NGC\,253 in the circular apertures {\it c} and {\it d} defined by Radovich \etal\ including 
our 870\,\um\ fluxes. These apertures have diameters of $5'$ centered well outside the nucleus on the
north-eastern ({\it c}) and south-western ({\it d}) part of the disk (see Fig.\,3 in Radovich \etal\ \cite{radovich01}). 
The 870\,\um\ fluxes derived from our LABOCA map are $1.4\pm1.5$ and $0.8\pm0.15$ Jy for aperture {\it c} and {\it d} respectively. 
Note that the additional apertures {\it e} and {\it f} in Radovich \etal, centered at the northern and southern outflow, 
are outside the field covered by our LABOCA observations. \\
As the optically thin approximation does not necessarily hold in the far/mid-IR regime, the dust emission was modeled using
\begin{equation}
S_{\nu}=\big( B_{\nu}(T_{\rm dust})-B_{\nu}(T_{\rm BG})\big)\big(1-e^{-\tau_{\nu}}\big)\,\Omega_{\rm s}\
\ , 
\end{equation}
where $B_{\nu}$ is the Planck function, $\tau_{\nu}$ the opacity and $\Omega_{\rm s}$ the source solid angle.
The dust optical depth was computed using:
\begin{equation}
\tau_{\nu}= \kappa_{\rm d}(\nu)\,M_{\rm
dust}/( D^2\,\Omega_{\rm s})  \ \ \ ,
\end{equation}
where $\kappa_{\rm d}$ is the dust absorption coefficient,
$M_{\rm dust}$ the dust mass and $D$ is the distance to the source. 
For the frequency dependence of the dust absorption coefficient we adopt
\begin{equation}
 \label{dustfreqdependence}
\kappa_{\rm d}(\nu)\ 
=\ 0.04\,(\nu/250\,{\rm GHz})^\beta\,
\end{equation}
with units of m$^{2}$ per kilogram of dust
(Kr\"ugel \& Siebenmorgen \cite{kruegel94}), and with $\beta= 2$ (Priddey
\& McMahon \cite{priddey01}).\\
\\
The observed dust SEDs for the two apertures are shown in
Fig.\,\ref{n253_dust} (left and right). Both SEDs require a fit with at least 3 temperature components to account for 
all observed fluxes between 870\,\um\ and 25\,\um. The dust SED for the north-eastern aperture is well described by a 17\,K and a 28\,K dust 
component for observations up to 60\,\um. The 25\,\um\ observations require higher dust temperatures and we 
have added, for illustration purposes, a 60\,K component to both SEDs. For the south-western aperture 
the 180\,\um\ flux is too high in comparison to our LABOCA flux so that both data points cannot be fit with a single dust 
temperature. The IRAS 100\,\um\ point, however, limits the temperature of 
the cold dust to $\le20$\,K and we in the following therefore adopt a dust temperature of 17\,K for the entire disk of NGC\,253.\\
{ We note that the disk SED for NGC\,253 (as well as for NGC4945 and Cen\,A) is also consistent with additional cold gas below
10\,K if the mass for the 17\,K dust component is reduced. From our SED fits we find that such cold gas could contribute about 30--40\% to
the observed 870\microns\ flux.}\\
We have also reanalyzed the dust SED towards the center of NGC\,253. At FIR wavelengths we used archival ISO 
Long-Wavelength Spectrometer (LWS) and Short-Wavelength Spectrometer (SWS) observations
which continously cover wavelengths between 30 and 197\,\um. The LABOCA flux in the $80''$ aperture of the ISO spectra is $7.5\pm0.8$\,Jy.
We used the CO(3--2) observations by Dumke \etal\ (\cite{dumke01}) to correct for the CO line emission in our 60\,GHz bandwidth. The CO(3--2)
line contributes 1.7\,Jy ($\sim 20\% $). In addition we have used the 1.3\,mm (Sakamoto \etal\ \cite{sakamoto06}), 
1.3\,mm (Kr\"ugel \etal\ \cite{kruegel90}), corrected for 25\% contribution from CO(2--1), (Mauersberger \etal\ \cite{mauers96}) 
and 350\,\um\ (Rieke \etal\ \cite{rieke73}) fluxes towards the center of NGC\,253.\\
The SED towards the center is well described by a two component dust model. The dust temperatures, however, depend
on the underlying source solid angle. If e.g. we assume that the true source size matches the deconvolved source size derived
above ($30''\times17''$) we get dust temperatures of 30 and 55\,K. Even for this large source size the dust optical depth 
is not negligible in the FIR regime ($\tau_{100\um}=0.31$). If we assume that the deconvolved source size has an area filling factor of only
10\%, as suggested from CO line observations at similar spatial resolution (Mauersberger \etal\ \cite{mauers96}), the 
FIR opacities increase to $\tau_{100\um}=3$ and the dust temperature for the colder component increases to 37\,K.
 In any case the mm/submm fluxes in the central $80''$ region of NGC\,253
are dominated by the warm central starburst region and do not show strong evidence for dust with temperatures below 30\,K.\\
We use our coldest dust temperature estimate in the disk and in the nucleus to calculate the total gas mass of NGC\,253 
from the optically thin 870\,\um\ emission: 
\begin{equation}
M_{\rm gas}= S_{870}\,{\rm D}^2\,\kappa^{-1}_{\rm gas\,870}\,\big(B_{\rm 870,T_{\rm dust}}-B_{\rm 870,T_{\rm BG}}\big)^{-1}.
\end{equation}
Using $\kappa_{\rm gas\,870}=0.0005\,{\rm m}^{2}\,{\rm kg}^{-1}$, which corresponds to a 870\,\um\ emissivity according to Eq.\,3 and a gas
to dust mass ratio of 150, we derive a total gas mass of $M_{\rm total}\approx2.1\times 10^{9}\,\msol$. 
The gas mass associated with the warm (35\,K) dust in the nuclear region is $M_{\rm center}\approx3.0\times 10^{8}\,\msol$.

\subsection{NGC\,4945}
\begin{figure*} 
\centering
\includegraphics[width=16.5cm,angle=0]{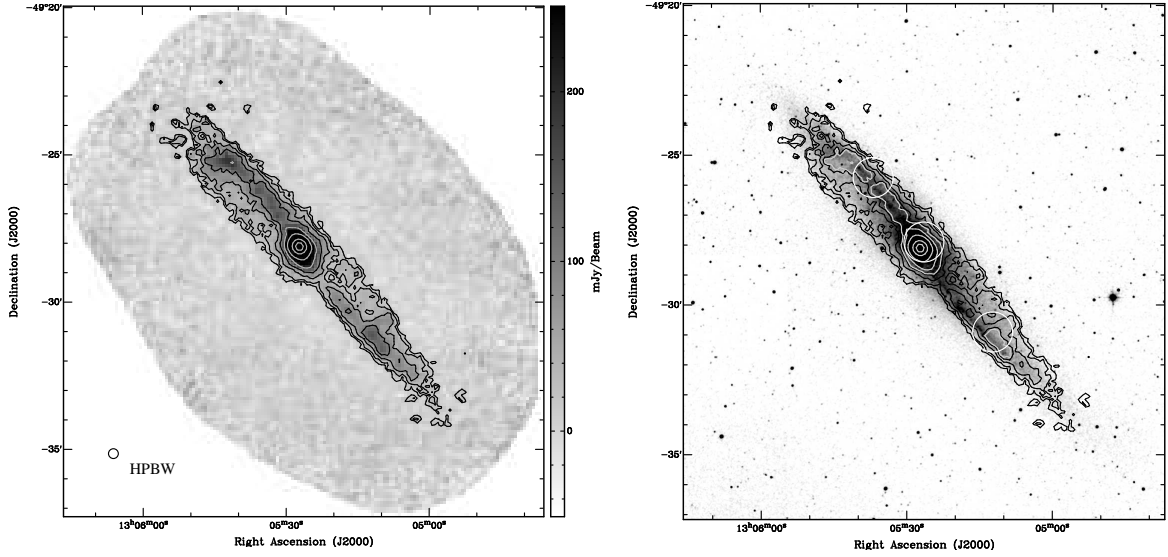} 
\caption{{\it Left:} LABOCA 870\,\um\ flux density map of NGC\,4945 smoothed to a spatial resolution of 20$''$. Contours are show at 0.015, 
0.03, 0.06, 0.1 (black) and 0.2, 0.5, 2.0, 4.0 Jy/beam (grey). {\it Right:} 870\,\um\ contours overlaid onto a 2MASS K-band image of NGC\,4945. 
Contour levels for 870\,\um\ are the same as in the left part of the figure { with an additional grey contour at 0.11 Jy/beam}. The circles show 
the apertures of the ISO LWS observations.}
\label{n4945_image} 
\end{figure*}

\begin{figure*} 
\centering
\includegraphics[width=16.5cm,angle=0]{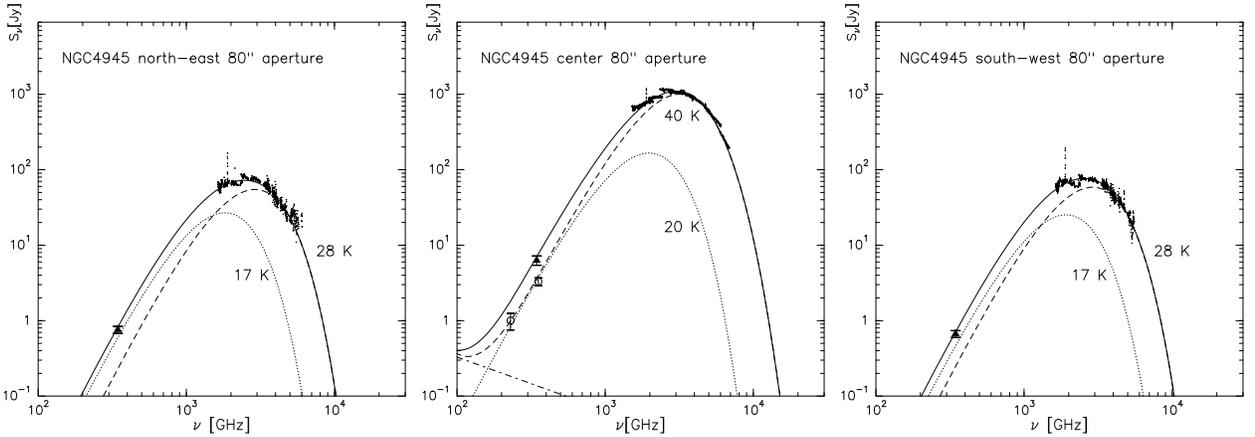} 
\caption{Spectral energy distributions towards the north-east (left), center (middle) and south-west (right) of NGC\,4945
measured in a 80$''$ aperture. The location of the apertures is shown in Fig.\ref{n4945_image} (right).
The mid-IR data are from archival ISO LWS observations. The open circles in the central plot represent the 1.2\,mm and 870\,\um\ fluxes 
associated with the 40\,K gas in the central $185\times95$\,pc (Chou \etal\ \cite{chou07}).
The dashed dotted line for the center of NGC\,4945 shows the non-thermal contribution of
the AGN to the spectrum.}
\label{n4945_seds} 
\end{figure*}

{\it 870\,\um\ morphology:} The LABOCA 870\,\um\ map of NGC\,4945 is shown in Fig.\,\ref{n4945_image} (left). The bright central starburst 
region has a deconvolved size of $19''\times15''$ (FWHM, $370\times280$\,pc at a distance of 3.8\,Mpc, Karachentsev \etal\ 
\cite{karachentsev07}) and an integrated flux of $7.2\pm0.8$\,Jy. The dust emission from the disk of NGC\,4945 follows closely the spiral arms 
visible in the K-band (see Fig\,\ref{n4945_image}, right). Weak 870\,\um\ emission is detected across the entire optical disk. 
The diffuse emission on the north-eastern remanent side of the galaxy is visible as strong dust absorption in optical B-band images.
The total 870\,\um\ flux detected in NGC\,4945 is $15.8\pm1.6$\,Jy.\\
{\it Dust temperature and mass:} For the determination of dust temperatures in the center of NGC\,4945 we have used archival ISO 
LWS data ranging from 
43 to 197\,\um. ISO also observed two additional positions in the NE and SW part of the disk. The ISO
pointings are shown as circles in Fig.\,\ref{n4945_image} (right). These two observations only cover wavelengths shorter than 183\,\um\ 
and show high noise below 50\,\um. The noisy part has been omitted in the analysis. The dust SEDs of the three regions are shown 
in Fig\,\ref{n4945_seds}. As NGC\,4945 contains an AGN (e.g. Madejski et al. \cite{madejski00}) we have estimated the contribution 
of non-thermal emission to the observed 870\,\um\ flux density in the center using radio observations between 0.4 and 8\,GHz. We find 
that the non-thermal contribution is negligible at 870\,\um\ ($\sim 0.13$\,Jy, see Fig\,\ref{n4945_seds} center).
The dust SEDs at all three positions are well described by a two component dust model. The dust SEDs in the NE and SW are very 
similar and both are fit by temperatures of 17\,K and 28\,K - similar to those derived above for the disk of NGC\,253. The 28\,K component 
in the disk of NGC\,4945 is more pronounced  than in NGC\,253 and dominates the peak of the SEDs. The mass, however, is dominated 
by the 17\,K component.\\
For the determination of the dust temperature in the central $80''$ region we have used the high-spatial resolution 1.3\,mm observations 
from the Submillimeter Array (SMA, Chou \etal\ \cite{chou07}) to disentangle emission from the warm, compact starburst region and the disk. 
As the 1.3\,mm continuum emission is more compact than the LABOCA beam ($10''\times5''$) we first fit a warm dust component to the integrated SMA 
flux, the flux in the central LABOCA beam (3.3 Jy, CO line corrected) and the FIR ISO spectrum using $\Omega_s=56\,{\rm arcsec}^2$, the 
source solid angle from the 1.3\,mm SMA continuum observations. 
This approach is justified as most of the FIR emission also arises from a region smaller than the LABOCA resolution 
(Brock \etal\ \cite{brock88}). The contribution from the CO(3--2) line to the LABOCA flux was estimated using the flux densities 
from Wang \etal\ (\cite{wang04}) and found to be of order 20\%. This yields a dust temperature of 40\,K for the central $10''\times5''$.
The observed 870\,\um\ flux (also corrected for 20\% line contribution) 
and the long ISO wavelengths for the full $80''$ aperture requires an additional cold dust component which is well described with a 
temperature of 20\,K. Similar to our results on NGC\,253, the FIR opacity in the nuclear region is high 
with $\tau_{100\um}=1.1$, in agreement with the lower limit from Brock \etal\ (\cite{brock88}).\\
With the temperatures of the coldest, mass dominating dust components in the disk (17\,K) and in the center (20\,K) we obtain a total
gas mass for NGC\,4945 of  $M_{\rm total}\approx4.2\times 10^{9}\,\msol$ of which $1.2\times 10^{9}\,\msol$ are located in the central
$1.4$\,kpc. The warm (40\,K) gas mass associated with the IR luminous central $185\times95$\,pc is $3.8\times 10^{8}\,\msol$.

\subsection{Cen\,A (NGC\,5128)}

\begin{figure*} 
\centering
\includegraphics[width=16.5cm,angle=0]{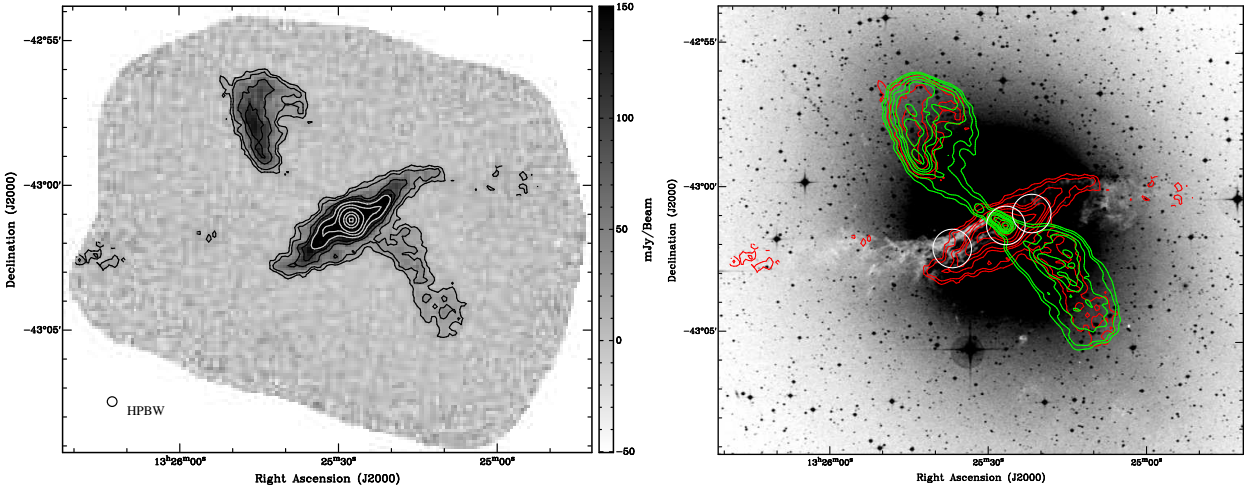} 
\caption{{\it Left:} LABOCA 870\,\um\ flux density map of Cen\,A  (NGC\,5128) smoothed to a spatial resolution of 20$''$. 
Contours are show at 0.015, 0.03, 0.07, 0.1 (black) and 0.15, 0.25, 0.5, 2.0, 4.0, 6.5 Jy/beam (grey). {\it Right:} 870\,\um\ 
contours (red) overlaid onto a B-band image of Cen\,A from the SDSS archive. Contour levels for 870\,\um\ are the same as 
in the left part of the figure. Green contours show the 1.4 GHz VLA radio continuum map (Burns \etal\ \cite{burns83}). Contours 
correspond to 0.3, 0.5, 1.0, 1.5, 2.0, 3.0 and 4.5 Jy/beam. The circles correspond to the apertures observed by ISO.}
\label{cena_image} 
\end{figure*}

\begin{figure*} 
\centering
\includegraphics[width=16.5cm,angle=0]{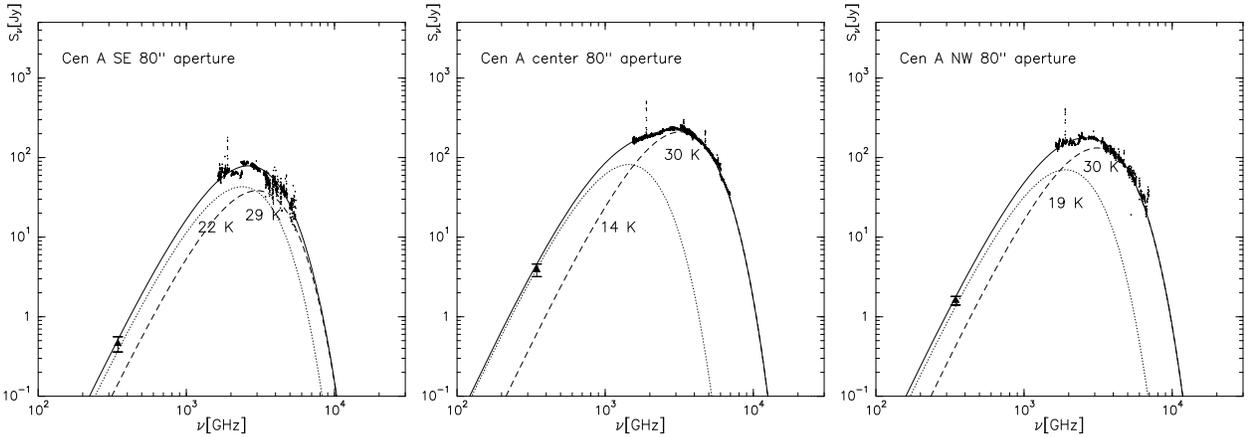} 
\caption{Spectral energy distributions towards the south-east (left), center (middle) and north-west (right) of NGC\,5128
measured in a 80$''$ aperture. The location of the apertures is shown in Fig.\ref{cena_image} (right).
The FIR-IR data are from archival ISO LWS observations.}
\label{cena_seds} 
\end{figure*}

Our LABOCA map of NGC\,5128 shows four distinct emission features: 1) the bright, central radio core which is blended with the 
central dust disk, 2) the northern and 3) southern radio lobes and 4) some faint emission associated with the optical dust absorption 
lanes in the outer parts of the galaxy. Emission from the lobes is detected for the first time at submm wavelengths and its 
distribution follows closely that of the radio emission at cm wavelengths (Condon \etal\ \cite{condon96}, see Fig\,\ref{cena_image}). 
They are discussed in more detail in  Sect.\,\ref{radiolobes}. 
The central dust disk has a pronounced S-shape and is very similar in structure to what has been observed in previous 
SCUBA and ISOCAM images at 850 and 450\,\um\ as well as at 15 and 7\,\um\ (Mirabel \etal\ \cite{mirabel99}, Leeuw \etal\ \cite{leeuw02}).\\
The total flux in our map is $18.0\pm1.7$\,Jy out of which $3.3\pm0.4$\,Jy are associated with the radio lobes. 
The unresolved central core has a flux density of $6.4\pm0.7$\,Jy which is lower than the 850\,\um\ flux of $8.1\pm0.8$\,Jy 
reported by Leeuw \etal\ (\cite{leeuw02}). This may suggest long term variability of the AGN in NGC\,5128 up to submm wavelengths.
Nevertheless calibration uncertainties or the effective reconstruction of the underlying extended dust disk may play a role too.\\
The dust temperatures in the disk were determined at three positions using our 870\,\um\ fluxes in conjunction with 
archival ISO-LWS spectra observed towards the nucleus and at equatorial offsets of $-53'',27''$ (NW) and $110'',-49''$ (SE) from 
the center (Unger \etal\ \cite{unger00}, Fig\,\ref{cena_image}, right). The south-eastern spectrum is very noisy for wavelengths 
below 55\,\um\ and we have 
omitted these data from our analysis. The LABOCA fluxes in the $80''$ ISO apertures are 0.5, 10.3 and 1.6 Jy for the SE, 
center and NW respectively. For the central aperture we have subtracted the 870\,\um\ flux of the central radio core (6.4 Jy, see above) 
to correct for the non-thermal contribution in this region. This, however, only
yields a proxy to the 870\,\um\ dust emission as we expect also a non-thermal contribution from the jets emerging from the 
central engine. This is evident from the shallow spectral index between 850 and 450\,\um\ in this region (Leeuw \etal\ \cite{leeuw02}).
The dust SEDs in the three regions are shown in Fig.\,\ref{cena_seds}.\\
All SEDs are well described by a two component dust model. For the warmer gas component we find in all three apertures a dust temperature
of $\sim30$\,K in agreement with the results from Unger \etal\ (\cite{unger00}). The cold component in the NE has a dust temperature of 19\,K
similar to the disk temperatures of NGC\,253 and NGC\,4945. In the SE our fit indicates that the cold gas may be slightly warmer with 22\,K.
The FIR spectrum in the SE, however, is quite noisy and the data is also consistent with 20\,K. The SED fit to the central region yields a 
dust temperature of only 14\,K - significantly lower than what was derived in the NW and SE. We note that this even holds if we have 
underestimated the non-thermal 870\,\um\ contribution: in order to be consistent with a 19\,K dust spectrum the 870\,\um\
emission in the central aperture (after correcting for the flux from the radio core) needs to be dominated by the non-thermal emission
from the jet which is inconsistent with the measured surface brightness from the southern radio lobe near the disk. 
The low temperature for the cold dust in the central region is consistent with the results from Leeuw \etal\ (\cite{leeuw02}) 
who derive 12\,K albeit in a larger aperture. Our analysis, however, does not confirm their decreasing dust temperature of the 
warm component with increasing distance from the center.\\
Using 14\,K for the the cold dust in the central aperture and 20\,K as for the rest of the thermal 870\,\um\ emission we derive a
total gas mass for NGC\,5128 of $M_{\rm total}\approx2.8\times 10^{9}\,\msol$ out of which $M_{\rm center}\approx1.6\times 10^{9}\,\msol$
are associated with the 14\,K gas in the central 1.4\,kpc ($80''$ aperture).

\begin{table*}
\caption[]{Observed and derived dust properties of our target galaxies.}
\label{dusttable}
            
\begin{tabular}{l c c c c c c c}
\hline
\noalign{\smallskip}
Target & D    & total observed Flux & $T_{\rm dust,disk}$ & $M_{\rm dust,disk}$ &  $T_{\rm dust,center}$ & $M_{\rm dust,center}$ 
& $M_{\rm gas,total}^{\rm a}$\\
       & [Mpc]& [Jy]       &[K]      &[$10^{6}$ \msol]  &[K]      &[$10^{6}$ \msol]   &[$10^{9}$ \msol]          \\
            \noalign{\smallskip}
            \hline
            \noalign{\smallskip}
NGC\,253& 2.5 & $17.6\pm1.8$ & 17/28 & 11.7 & 35/55 &2.0 & 2.1\\ 
NGC\,4945&3.8 & $15.8\pm1.6$ & 17/28 & 19.8 & 20/40 &5.7 & 4.2\\
NGC\,5128&3.5 & $18.0\pm1.7^{b}$ & 20/30 & 8.0 & 14/30 &10.8 & 2.8\\
\end{tabular}
\begin{list}{}{}
\item[a]assuming a gas-to-dust mass ratio of 150  
\item[b]non thermal contribution:  6.4 Jy (nucleus), 3.3 Jy (radio lobes)
\end{list}
\end{table*}

\section{Discussion} 

\subsection{Dust temperatures}
Our dust temperature estimate for the coldest, mass carrying dust component of 17\,K in the disk of NGC\,253 is in excellent agreement 
with estimates based on lower resolution ISOPHOT and IRAS data (Melo \etal\ \cite{melo02}, Radovich \etal\ \cite{radovich01}). 
The dust SEDs in the disks of 
NGC\,4945 as well as NGC\,5128 show similarly low dust temperatures, which suggests that the temperatures of the dust in the disks of 
all three active galaxies are comparable to those found in the Milky Way (MW, e.g. Cox \& Mezger \cite{cox89}). None of our target shows 
evidence for dust with temperatures below 10\,K. 
This finding is also in agreement with results based on SCUBA and Spitzer observations of the SINGS galaxy sample 
(Draine \etal\ \cite{draine07}). { Our dust SEDs, however, do not rule out the presence of such cold gas. Its effect
on the total dust/gas mass estimate, on the other hand, is not dramatic because a cold (e.g. a 8\,K) dust component in the disks cannot 
contribute to more than $\sim$40\% of the observed 870\microns\ flux in order to be consistent with the overall SED shape.}\\
For the nuclear starburst regions our analysis suggests that NGC\,253 is free of cold (T$<$30\,K) dust while the 
central SEDs for NGC\,4945 and NGC\,5128 both show a clear signature of dust at temperatures comparable to those 
found in the disks. This, however, does not necessarily imply that the nuclear starburst itself contains
cold gas because the 80$''$ aperture analyzed here is still too large to separate spatially the central active region from the surrounding
disk. For NGC\,5128 the situation is further complicated by the strong non-thermal contribution from the central AGN and { the jets} to the 
870\,\um\ flux, which makes it difficult to determine the submm flux related to thermal dust emission alone. Therefore, 
the suggested decrease of the cold { dust component's temperature} from the disk towards the center in NGC\,5128 may simply reflect an imperfect
separation between the thermal and non-thermal flux contributions.

\subsection{Gas masses}
\begin{figure} 
\centering
\includegraphics[width=8.5cm,angle=0]{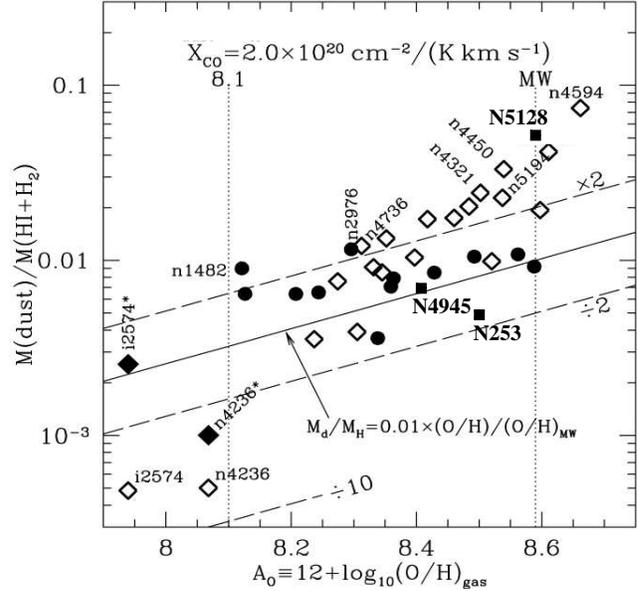} 
\caption{Dust-to-gas mass ratio vs. characteristic oxygen abundance plot for the SINGS galaxies adopted from Draine \etal\ (\cite{draine07}, their Fig.\,16).
Our target galaxies are shown as filled squares. The CO intensities have been converted to \hh\ masses using a conversion factor of 
$2.0\times10^{20}\,{\rm cm}^{-2}\,({\rm K\,\kms})^{-1}$.}
\label{dust_metall} 
\end{figure}

Due to the large areas our target galaxies project on the sky, only few mm/submm observations have been published so far which allow to 
determine the total molecular gas content across the entire optical disks. { For NGC\,253, Houghton \etal\ (\cite{houghton97})
have presented CO(1--0) observations having a similar extent} than our LABOCA maps. They find a total molecular gas mass 
of $2.4\times 10^{9}\,\msol$ using a CO conversion factor of $3\times10^{20}\,{\rm cm}^{-2}\,({\rm K\,\kms})^{-1}$. This value
includes the contribution of heavier elements.  The \hi\ mass of NGC\,253 is $1.8\times 10^{9}\,\msol$ (Koribalski \etal\ \cite{kori04}) 
which yields a total mass of $4.2\times 10^{9}\,\msol$, a factor of 2 higher than our estimate based on the dust continuum. This may argue for a 
smaller CO conversion factor or a higher gas-to-dust mass ratio. { Several studies have suggested that the average 
gas-to-dust mass ratio of a galaxy is a function of its enrichment (e.g. Draine \etal\ \cite{draine07}, Engelbracht \etal\ \cite{engel08}). Indeed
the characteristic oxygen abundance (a measure for the average metallicity of a galaxy) in NGC\,253 is lower by  $\sim 0.1$\,dex  than the MW's 
metallicity (Pilyugin \etal\ \cite{pil04}). But such a small difference in metallicity is not expected to cause the
 higher gas-to-dust mass ratio in NGC\,253 illustrated in Fig.\,\ref{dust_metall} where we show the location of our target galaxies on 
the dust-to-gas mass ratio vs. characteristic oxygen abundance plot for the SINGS galaxies (Draine \etal\ \cite{draine07}, their Fig\,16).
NGC\,253 lies well within the factor 2 scatter band on this metallicity relation observed for the SINGS galaxies. Given the simplifying 
assumptions made in our mass estimate (e.g. the constant dust absorption coefficient across the disk of NGC\,253 and the use of constant 
CO conversion factor), the mass estimates for the dust continuum and the CO and \hi\ are in reasonable agreement.}\\


\noindent CO(1--0) observations covering the optical disk of NGC\,4945 have been published by Dahlem \etal\ (\cite{dahlem93}). 
They give an \hh\ mass of $2.2\times 10^{9}\,\msol$ (scaled to D\,=\,3.8\,Mpc and using a CO conversion factor of $2\times10^{20}\,{\rm cm}^{-2}\,
({\rm K\,\kms})^{-1}$). Correcting for heavier elements and taking  the \hi\ mass of $1\times 10^{9}\,\msol$ (Ott \etal\, \cite{ott01}, 
scaled to D\,=\,3.8\,Mpc) into account yields a total gas mass of $4\times 10^{9}\,\msol$, in good agreement with our results. { Our dust mass
is also in agreement with the expected metallicity dependence for MW like dust (see Fig\,\ref{dust_metall}). For the determination of the characteristic
oxygen abundance we have used here the O/H --$M_{B}$ relation for spiral galaxies from Pilyugin \etal\ (\cite{pil04}).}\\
\\
For NGC\,5128 the situation is different as our estimated gas mass based on the dust emission exceeds those based on CO 
and \hi\ by more than a factor of 4: Eckhart \etal\ (\cite{eckhart90}) derive a total molecular 
gas mass based on CO of $2.7\times 10^{8}\,\msol$ (scaled to D\,=\,3.5\,Mpc). The \hi\ mass accounts for  
$2.5\times 10^{8}\,\msol$ (Gardner \& Whiteoak \cite{gardner76}, scaled to D\,=\,3.5\,Mpc). Correcting for 
heavier elements this yields a total gas mass of  $6\times 10^{8}\,\msol$ compared to our estimate of $2.8\times 10^{9}\,\msol$. 
We note that this discrepancy still holds if we would assume that the cold dust towards the center has a temperature of 20\,K 
instead of the 14\,K derived  from our SED fitting (see discussion above).\\
Based on SCUBA 850 and 450\,\um\ observations, Leeuw \etal\ (\cite{leeuw02}) find a dust mass of $2.2\times 10^{6}\,\msol$ for the 
central 4.5\,kpc which translates into a gas mass of $3.3\times 10^{8}\,\msol$ assuming a gas-to-dust mass ratio of 150 -- a factor 
of two lower than the estimate based on CO and a factor of 8 lower than our estimate. This discrepancy is difficult to understand 
because our assumed dust absorption coefficient, $\kappa_\nu$, differs at most by $\sim30\%$ from the values used 
by  Leeuw \etal\ and our LABOCA flux is only slightly higher than the SCUBA 850\,\um\ flux 
(7.8 Jy compared to 6.4 Jy adding the inner and outer disk as defined in Leeuw \etal). Furthermore, our dust temperature is 
somewhat higher than the 12\,K derived by Leeuw \etal, so that we would expect only a small difference between the two dust mass estimates. 
Using the dust temperatures, emissivities and 850\,\um\ fluxes given in  Leeuw \etal\ we indeed estimate dust and gas 
masses of $1.1\times 10^{7}\,\msol$ and  $1.7\times 10^{9}\,\msol$ for NGC\,5128 in better agreement with our results. 
Our higher dust masses are also in agreement with the results of Mirabel \etal\ (\cite{mirabel99}).
Their 850\,\um\ fluxes, however, are about a factor of 10 higher than those published by Leeuw \etal\ (\cite{leeuw02}) and from our 
LABOCA maps and their dust mass should therefore be treated with some caution. 
Independent estimates of the dust mass in NGC\,5128 have been derived by Block \& Sauvage (\cite{block00}) using the 
V-15\,\um\ ratio which results in $M_{\rm d}=2.3\times10^{6}\,\msol$ (scaled to D\,=\,3.5\,Mpc). This study, however, 
only addresses the dust content of the central 90$''$ and may miss diffuse dust which could significantly increase the 
estimated dust mass. \\
Considering these uncertainties in the dust mass in Cen\,A it is impossible to obtain strong conclusions by comparing 
the gas mass derived from CO and the dust continuum. We note, however, that the molecular gas mass derived by  Eckart \etal\ 
(\cite{eckhart90}) is based on a standard galactic conversion factor and therefore unlikely to underestimate the 
gas mass by more than a factor of 2. Therefore, the high dust mass derived from our LABOCA data may suggest that the dust 
properties or the gas-to-dust mass ratio in NGC\,5128 are different from those in NGC\,253 and NGC\,4945.\\
{ To our knowledge no oxygen abundance estimates across the disk of NGC\,5128 has been published which would allow 
secure derivation of the characteristic oxygen abundance to check for metallicity effects as a possible reason for the 
low gas-to-dust mass ratio. Estimates based on the globular cluster systems of NGC5128 indicate metalicities similar 
to or somewhat higher than their MW counterparts (see Israel \etal\ \cite{israel98} and references therein). Using, 
as for NGC\,4945, $M_B$ as indicator for O/H leads to a similar result (12+log(O/H)\,=\,--8.5) although it is not clear if 
this relation holds for elliptical galaxies. We therefore assume a characteristic oxygen abundance similar to the MW which 
places NGC5128 well above the expected metallicity dependence in Fig.\,\ref{dust_metall}.}

\subsection{CO intensity vs dust column density in NGC\,253}
Among our target galaxies the CO emission in NGC\,253 has been studied in most detail and large scale CO maps exist in the literature with 
similar spatial resolution than our LABOCA data. Sorai \etal\ (\cite{sorai00}) imaged large parts of the 
optical disk in CO(1--0) at a spatial resolution of $16''$ using the  Nobeyama 45m telescope. { To investigate variations of 
the mass tracing capabilities of CO in comparison to the dust continuum we have compared their integrated CO intensities to 
the \hh\ column densities derived from our LABOCA maps within the central 6\,Kpc of NGC\,253 
taking the neutral gas mass fraction measured from \hi\ into account. 
The gas mass (\hi\ \& \hh) distribution from the dust was computed using a dust temperature distribution as suggested from 
our dust SED analysis: 35\,K within the central  $30''\times 16''$, 19\,K out to a radial distance of $40''$ (ISO aperture) and 17\,K for 
emission further out and a constant gas-to-dust mass ratio of 150. This gas mass distribution was corrected for \hi\ using the 
radial dependence of molecular gas mass fraction from Sorai \etal\ (\cite{sorai00}, their Fig.\,9\,a) and finally converted to a 
\hh\ column density map.}

\begin{figure*} 
\centering
\includegraphics[width=16.0cm,angle=0]{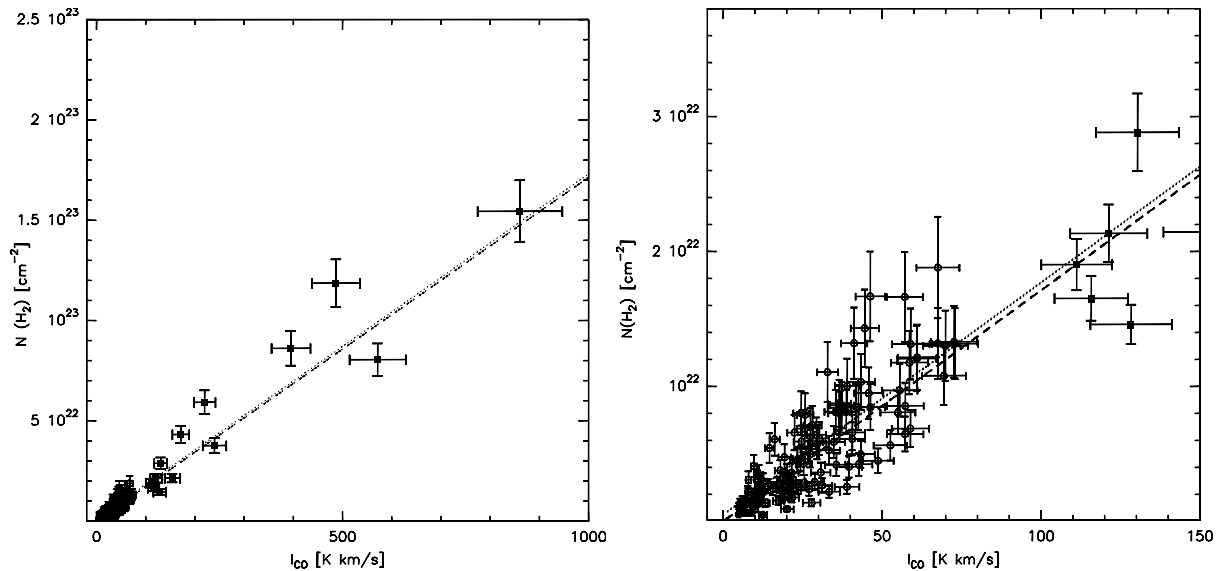} 
\caption{Observed integrated CO(1--0) intensities (Sorai \etal\ \cite{sorai00}) compared to the \hh\ gas column densities derived 
from our dust continuum analysis in NGC\,253 { corrected for the \hi\ contribution}. Filled squares show positions towards 
the central starburst region, open circles positions in the bar/disk of NGC\,253. The dashed line is a linear fit to data in the 
starburst region only, the dotted line a fit to all other data points. The right part of the figure is a zoom to the full diagram 
shown on the left.}
\label{xco} 
\end{figure*}

\noindent The comparison was done for { 20$''$ sized pixels after smoothing both data 
sets to that resolution} and is shown in Fig.\,\ref{xco}. From this figure there is no obvious difference 
between the CO conversion factor in the central starburst region and positions in the bar/disk region of NGC\,253.
A linear fit yields conversion { factors of $1.7\times10^{20}\,{\rm cm}^{-2}\,({\rm K\,\kms})^{-1}$ 
(2.7\,\msol\,(K\,\kms\,pc$^2$)$^{-1}$) in both regions with errors of 3\% and 7\% for the central
and the bar/disk region}. { The larger error for the slope outside the nuclear starburst region is mainly due 
to the smaller dynamic range of the data points. The result is little dependent on the neutral gas correction. 
For example, if we assume that all gas is in molecular form and ignore the \hi\ (which maximizes the conversion factor),
we find the same value for the nuclear region and an increase of only 50\% ($2.5\times10^{20}\,{\rm cm}^{-2}\,({\rm K\,\kms})^{-1}$)
for the bar/disk region.}

\noindent This result is surprising at the first 
glance, as several detailed CO studies have suggested that the CO conversion factor is { several times} lower 
in the central starburst region { (e.g. Mauersberger \etal\ \cite{mauers96}, Harrison \etal\ \cite{harrison99} for 
the center of NGC\,253)} than the galactic conversion factor, which { in turn} is expected to be a good approximation 
for the regions outside the starburst. { In contrast Fig.\,\ref{xco} hints at a constant conversion factor over the entire 
galaxy, including its nucleus. One must note, however, that in the strict sense, Fig.~\ref{xco} implies a constant ratio only 
for the measured CO/dust masses (or fluxes) across the galaxy, while the conversion of these into total gas mass may both also 
depend on other characteristics. Specifically, our result is consistent with the possibility of both the gas/CO and the gas/dust 
ratios varying similarly with metalicities. For example, Mauersberger \etal\ (\cite{mauers96}) assumed a metallicity 
of 3 times the solar value for their determination of the CO conversion factor in NGC\,253. Matching this, a gas-to-dust mass 
ratio that scales inversely with metalicity, i.e.~50 (if a galactic value of 150 is assumed) for the nucleus of NGC\,253, is 
implied by our results. The apparent constancy of the CO/dust ratio across NGC253 also suggests that changes in the gas/CO 
conversion are unlikely to be dominated by the variation of CO excitation, since that would be apparent in deviations from the 
strict linear relation, which we do not see. }

\subsection{Radio lobes in Cen\,A\label{radiolobes}}

\begin{figure} 
\centering
\includegraphics[width=9.0cm,angle=0]{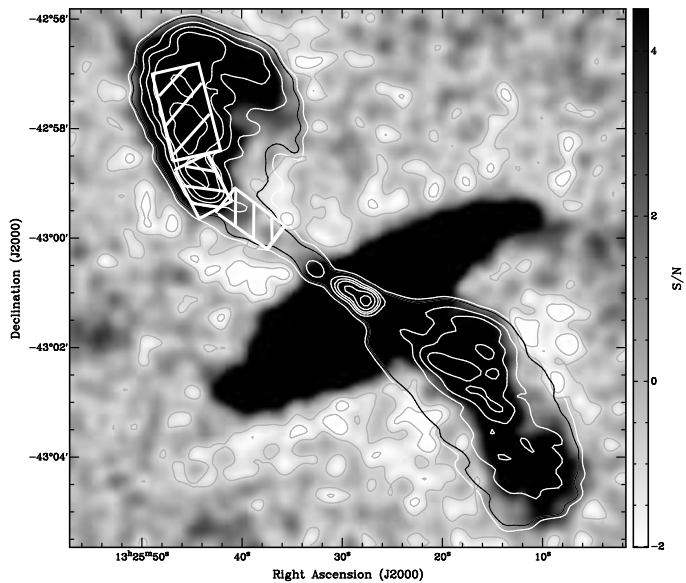} 
\caption{High contrast representation of the signal to noise map of our 870\,\um\ observations towards NGC\,5128 smoothed to an effective
spatial resolution of 25$''$ (beam smoothed). { The negative parts of the image are highlighted by two grey contours with signal to 
noise ratios of $-1$ and $-2$.} 
This presentation also shows the faint submm emission from the northern radio jet. The white and black contours are the 1.4\,GHz VLA map from 
Condon \etal\ (\cite{condon96}). The { white} boxes indicate those regions which have been used to analyze the spectral index.}
\label{cena_lobes} 
\end{figure}

\begin{figure} 
\centering
\includegraphics[width=8.5cm,angle=0]{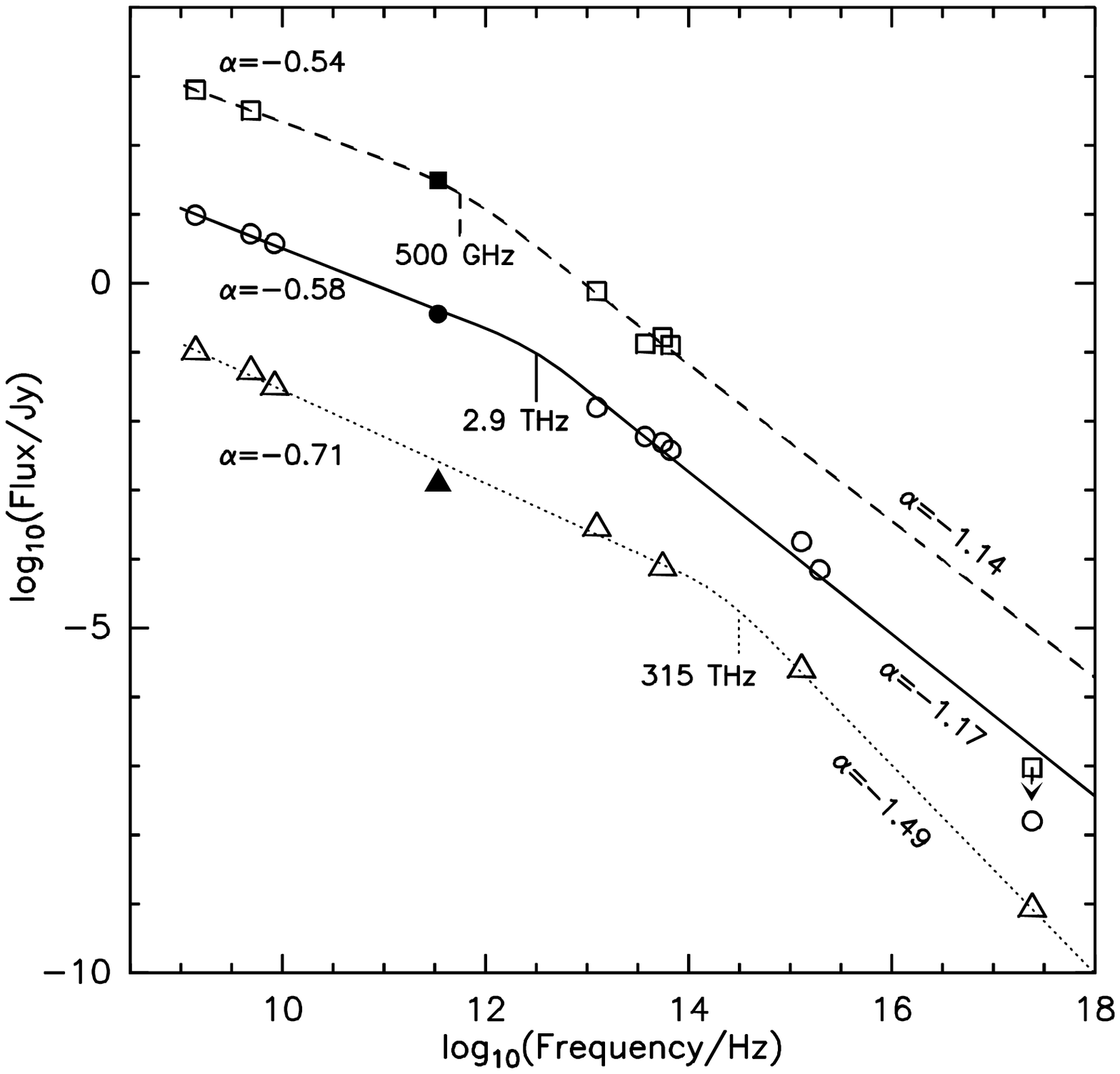} 
\caption{Synchrotron spectra from radio to X-rays in three apertures in the northern radio lobe. The apertures are shown in Fig.\,\ref{cena_lobes}
and correspond to the "outer" (shown as dashed line and square symbols), "middle" (solid, circles) and "inner" (dotted, triangles) region 
defined in Hardcastle, Kraft \& Worrall \cite{hardcastle06}. The filled symbols represent
the 870\,\um\ flux densities. For all other data see Table\,1 in Hardcastle, Kraft \& Worrall (\cite{hardcastle06}). The lines show 
broken power law fits to the data. The presentation for the outer/inner region have been up/down scaled by 1.5 dex.}
\label{power} 
\end{figure}

The 870\,\um\ emission associated with the radio lobes is very similar in morphology to the 1.4GHz continuum 
emission (Fig.\,\ref{cena_image}\,right \& Fig.\,\ref{cena_lobes}). As the synchrotron emission from the northern radio lobe has also been 
detected at 15--4.5\,\um\, in the UV and at X-ray energies (Brookes \etal\ \cite{brookes06}, Hardcastle, Kraft \& Worrall \cite{hardcastle06}) 
it is clear that the 870\,\um\ emission is synchrotron in nature and not associated with dust driven out from the disk by the radio jet.
The southern radio lobe is detected for the first time at wavelengths shorter than 3.5\,cm. We show a high
contrast signal to noise map of our LABOCA observations in Fig.\,\ref{cena_lobes}. From this image it is evident that
the northern radio jet is also detected at 870\,\um. The jet, however, is located in a low level negative bowl surrounding 
the strong central emission. Such negative artifacts are unavoidable byproducts of the skynoise removal which results in 
the filtering of low spatial frequencies. As a consequence the true 870\,\um\ flux of the jet is difficult
to determine. The 870\,\um\ flux of the northern and southern lobes are 2.5 and 0.9\,Jy respectively.\\
The variation of the spectral index of the northern lobe with wavelength has recently been investigated by 
Hardcastle, Kraft \& Worrall (\cite{hardcastle06}) and Brookes \etal\ (\cite{brookes06}). Based on 
{ Spitzer/IRAC}, GALEX and Chandra data Hardcastle \etal\ fit the synchrotron emission by a broken power law. 
Their model with a break frequency at $\sim 300\,\rm{GHz}$, however, underestimates the 
observed UV flux and also provides a poor fit to the observed slope between the IRAC bands. We have included our LABOCA fluxes
in the analysis of the spectral index in the three regions in the northern lobe defined by Hardcastle \etal\ (see Fig.\,\ref{cena_lobes}). 
These regions cover the radio jet (inner), the southern part of the radio lobe (middle) and the northern part of the lobe 
(outer region). The LABOCA flux within the three regions is $40\pm20$\,mJy, $365\pm40$\,mJy and $1.0\pm0.1$\,Jy for the inner, 
middle and outer region respectively. The synchrotron spectra of the three regions are shown in Fig.\,\ref{power}. We fit the radio 
to X-ray data of all three regions  with a non-standard ($\Delta\alpha\not=0.5$), broken power law. For the inner region all data 
can well be fit with a break frequency of  $\rm{log}_{10}(\nu_{\rm b/Hz,inner})=14.5\pm0.65$ (315\,THz) and a change of the power index 
of $\Delta\alpha_{\rm inner}=0.8\pm0.3$. { Our LABOCA flux falls somewhat below the expected value from this model (68\,mJy) which 
we attribute to the difficulties to determine the 870\,\microns\ flux density in this particular region as described above. 
Because the break frequency in the inner region falls into the NIR regime this data point, however, is not critical for the fitting.}
For the middle and the outer regions the break frequency shifts, as expected, to lower 
frequency yielding $\rm{log}_{10}(\nu_{\rm b/Hz,middle})=12.5\pm0.3$ (2.9\,THz) and 
$\rm{log}_{10}(\nu_{\rm b/Hz,outer})=11.7\pm0.6$ (500\,GHz). The change of the power index for both regions is close to that 
predicted by a continuous injection model (KP model) with $\Delta\alpha_{\rm middle}=0.6\pm0.1$ and $\Delta\alpha_{\rm outer}=0.6\pm0.15$. 
For both regions the broken power law model, however, overpredicts the X-ray flux, which indicates that the energy distribution 
of the electrons is more complex than a distribution yielding a simple broken power-law. A detailed modeling of the energy 
distribution is beyond the scope of this paper but we note, that also a single injection (SI) model with its exponential
decrease at high energies does not provide a good fit to the data. From Fig.\,\ref{power} it can be seen that a model intermediate between
the SI and KP, e.g. a broken power law with a high energy exponential decrease, presumably gives a better description
of the observed fluxes between the radio and X-ray.\\
For the inner jet region, where the broken power law fits all observed fluxes, we can also estimate the speed of the electrons 
using our fit to the break frequency: assuming a magnetic field strength of 3\,nT, the minimum energy value 
(Brookes \etal\ \cite{brookes06}), we get an age of 
the electrons of $\sim\,17000$\,yrs which yields a speed of 0.5\,{\it c} using the measured projected distance of 2.6\,kpc from the 
nucleus and a jet view angle of $\sim20^{\circ}$ (Hardcastle \etal\ \cite{hardcastle03}). This speed is consistent with the 
electron speed derived close to the central source. (Hardcastle \etal\ \cite{hardcastle03}).

\section{Summary}

New large scale LABOCA 870\,\um\ continuum images are presented for the starburst galaxies NGC\,253, NGC\,4945 and the nearest giant 
elliptical radio galaxy Cen\,A (NGC\,5128). Our LABOCA images reveal for the first time the distribution of cold dust across the 
entire optical disks in NGC\,253 and NGC\,4945  out to a radial distance of 7.5 kpc. In NGC\,5128 the 870\,\um\ image shows not only 
the thermal emission from the central molecular gas disk and the optical absorption features but also the synchrotron emission 
from the central radio source, the northern radio jet and the inner radio lobes north and south of the disk. We have used the 870\,\um\ 
emission in conjunction with ISO-LWS, IRAS and long wavelengths radio data to analyze the thermal dust emission in all three galaxies as 
well as the synchrotron emission in NGC\,5128. Our main findings are summarized as follows:

\noindent Similar to the dust emission in the disk of the Milky Way the thermal dust emission at 870\,\um\  in the disks of NGC\,253, 
NGC\,4945 and NGC\,5128 is dominated on large scales by cold dust with temperatures of 17-20\,K. We do not find evidence for 
significant flux from even colder gas at temperatures below 10\,K

\noindent In the active centers, the lowest dust temperature appears to be at 30-40\,K. The dust SEDs of the central regions of NGC\,4945 
and NGC\,5128 do, however, show the presence of gas at temperatures comparable to that of cold gas in the disks. From our analysis we cannot 
rule out that this gas is associated with the starburst region itself, but it appears more likely that this gas is also associated 
with the disks. Higher spatial resolution FIR data will be required so solve this issue.

\noindent Using a gas-to-dust mass ratio of 150 we derive total gas mass of 2.1, 4.2 and $2.8\times10^{9}\,\msol$ for NGC\,253, NGC\,4945 
and NGC\,5128. While these estimates lead to consistent results compared to CO and \hi\ measurements for NGC\,253 and NGC\,4945, our 
mass estimate for NGC\,5128 exceeds { published gas mass estimates by a factor of 4}. This could suggest that the dust properties in NGC\,5128 are 
different from those in other active galaxies. The strong contribution of non-thermal emission at 870\,\um\ in NGC\,5128, however, 
complicates the analysis.

\noindent A detailed comparison between the gas masses based on the dust continuum and the observed CO(1--0) intensity in NGC\,253 shows 
no changes of the CO conversion factor towards the nuclear region if the gas-to-dust mass ratio is assumed to be constant across the 
galaxy. This suggests that at least for NGC\,253 changes of the CO conversion factor are mainly driven by a metallicity gradient 
and only to a lesser degree by variations of the CO excitation.

\noindent The contribution of the non-thermal emission processes in NGC\,5128 to the observed 870\,\um\ flux accounts for almost 50\%. 
The synchrotron spectrum (radio to the UV/X-rays) in the northern radio lobe is well described by a non-standard broken 
power-law spectrum. The break frequency is a function of the distance from the central radio source as expected for aging electrons. 
The break frequency of the synchrotron spectrum in the northern radio jet, 2.6\,kpc from the center, suggests 
an outflow speed of 0.5\,{\it c}, consistent with the speed derived very close to the nucleus.

\begin{acknowledgements}
We thank the APEX staff for their support during the LABOCA commissioning and observing run. We also thank K. Sorai for providing us 
with their Nobeyama 45m CO(1--0) data as well as E. Kr\"ugel for many fruitful discussions.
\end{acknowledgements}

\end{document}